# A Topos Model for Loop Quantum Gravity

**Tore Dahlen**

Department of Mathematics

The Faculty of Mathematics and Natural Sciences

University of Oslo

**Abstract**

One of the main motivations behind so-called topos physics, as developed by Chris Isham and Andreas Döring [4-7], is to provide a framework for new theories of quantum gravity. In this article we do not search for such theories, but ask instead how one of the known candidates for a final theory, loop quantum gravity (LQG), fits into the topos-theoretical approach. In the construction to follow, we apply the 'Bohrification' method developed by Heunen, Landsman and Spitters [10, 11] to the C*-algebra version of LQG introduced by Christian Fleischhack [9]. We then bring together LQG results and methods from topos physics in a proof of the non-sobriety of the external state space $\Sigma$ of the Bohrified LQG theory, and show that the construction obeys the standard requirements of diffeomorphism and gauge invariance.

## 1. The Topos-theoretical Approach to Quantum Physics

### 1.1. The Neo-realism of Döring and Isham

In a series of articles [4-7], Chris Isham and Andreas Döring have proposed a set of new models for quantum physics, dubbed as *neo-realism*. Neo-realism is conceived as an alternative to the well-known *Copenhagen interpretation*, which introduces a separation of the measurement process for a physical magnitude into two components, a quantum system *S* and a classical observer *V*. The possible states of *S* are wave functions $\Psi$ from a configuration space into the set of complex numbers, whereas the observer *V* always registers a real value as the outcome of his experiment. In Copenhagen terminology, the wave function "collapses" onto the registered value with probability

$$P(r) = |\langle r \mid \Psi \rangle|^2.$$

The physically meaningful (real) value *r* is not a value of the physical quantity before the measurement is made. The interpretation breaks down for closed systems where no "outside" observer is to be found, such as quantum cosmology. In the topos scheme suggested by Isham and Döring, physical quantities *does* have a value independent of any observer *V*. The scheme relies on non-standard representations of the states and quantity values of physics. It also turns out that a new, intuitionistic quantum logic supplants the familiar non-distributive logic of Birkhoff and van Neumann [2]. (It should be noted that the choice of the tag "neo-realism" would be protested by philosophers and logicians, such as Michael Dummett, who regard acceptance of the law of excluded middle as the hallmark of philosophical *realism* (e.g. [8], p. 130ff).)

In this subsection, we only give a brief outline of the topos scheme introduced by Döring and Isham. In order to appreciate the scope of the models, it is necessary to read the original articles. Also, notions from topos theory are used without explanation or comment (see [14] for a proper introduction to this field.)

Following earlier work by Isham and Butterfield, Döring and Isham [5] start their approach to quantum systems by assuming that the physical quantities *A* of a system *S* are represented by self-adjoint operators $\hat{A}$ in the non-commutative von Neumann algebra, $\mathcal{B}(\mathcal{H})$, of all bounded operators on the separable Hilbert space $\mathcal{H}$ of the states of *S*. The unital, commutative subalgebras *V* of $\mathcal{B}(\mathcal{H})$ are then considered as *classical contexts* or *perspectives* on the system *S*, and the *context category* $\mathcal{V}(\mathcal{H})$ is defined with $\text{Ob}(\mathcal{V}(\mathcal{H}))$ as the set of contexts *V* and $\text{Hom}(\mathcal{V}(\mathcal{H}))$ given by the inclusions $i_{V'V} : V' \to V$.

In general, a context *V* will exclude many operators. But, in a certain sense, excluded projection operators $\hat{P}$ still have "proxys" in *V*. For note that $\hat{P}$, even if not present in the context *V*, may be approximated by the set (where $\mathcal{P}(V)$ is the complete lattice of projections in *V*, and the ordering $\succsim$ is defined as $\hat{Q} \succsim \hat{P}$ if and only if $\text{Im}\hat{P} \subseteq \text{Im}\hat{Q}$, or, equivalently, $\hat{P}\hat{Q} = \hat{P}$)

$$\delta(\hat{P})_V := \bigwedge \{\hat{Q} \in \mathcal{P}(V) \,|\, \hat{Q} \succsim \hat{P}\}. \tag{1}$$

Truth values may now be assigned the projectors in each context *V* by using the Gelfand spectrum $\Sigma_V$. This is the set

$$\Sigma_V := \{\lambda : V \to \mathbb{C} \,|\, \lambda \text{ is a positive multiplicative linear functional of norm 1}\}. \tag{2}$$

When $\hat{P}$ is a projection, the value $\lambda(\hat{P})$ is either 0 *(false)* or 1 *(true)*, so $\lambda$ behaves like a "local" state for *V*: it answers "yes" or "no" to the "question" $\hat{P}$. The construction of the state object, the representation of the physical state space in the topos scheme, may now be undertaken. The *state object* (or *spectal presheaf*) $\underline{\Sigma}$ is the element in the class of objects of the topos of presheaves over the context category $\mathcal{V}(\mathcal{H})$,

$$\tau := \mathbf{Sets}^{\mathcal{V}(\mathcal{H})^{\text{op}}},$$

such that $\underline{\Sigma}_V := \Sigma_V$ and, for morphisms $i_{V'V} : V' \to V$ in $\mathcal{V}(\mathcal{H})$, $\underline{\Sigma}(i_{V'V}) : \underline{\Sigma}_V \to \underline{\Sigma}_{V'}$ is defined by $\underline{\Sigma}(i_{V'V})(\lambda) := \lambda \,|_{V'}$ (the restriction of $\lambda : V \to \mathbb{C}$ to $V' \subseteq V$).

As a substitute for the notion of a state (that is, a global element of the state space, the existence of which is excluded by the topos version of the Kochen-Specker theorem of quantum mechanics), Döring and Isham define a *clopen subobject* $\underline{S}$ of $\underline{\Sigma}$ as a subfunctor of $\underline{\Sigma}$ (in the standard sense) such that the set $\underline{S}_V$ is both open and closed as a subset of the compact Hausdorff space $\underline{\Sigma}_V$ (with weak*-topology).

There is now, from Gelfand spectral theory, a lattice isomorphism between the lattice

$\mathcal{P}(V)$ of projections in $V$ and the lattice of closed and open *subsets*, $\text{Sub}_{\text{cl}}(\underline{\Sigma}_V)$, of $\underline{\Sigma}_V$:

$$\alpha : \mathcal{P}(V) \to \text{Sub}_{\text{cl}}(\underline{\Sigma}_V) \text{ where } \alpha(\hat{P}) := \{\lambda \in \underline{\Sigma}_V \mid \lambda(\hat{P}) = 1\} \equiv S_{\hat{P}}. \tag{3}$$

Both $\mathcal{P}(V)$ and $\text{Sub}_{\text{cl}}(\underline{\Sigma}_V)$ are Boolean algebras, so the law of excluded third holds ([14], p. 55). The commutative algebras $V$ are classical contexts within the theory, so it is proper that these lattices are Boolean. Certainly, extraordinary logic is the last thing we would expect to find when we are engaged in experimental physics. Extending this construction to the total context category $\mathcal{V}(\mathcal{H})$, Döring and Isham ([4], th. 2.4) prove that, for each projection $\hat{P} \in \mathcal{B}(\mathcal{H})$, there is a clopen sub-object $\underline{S}_{\hat{P}}$ of the spectral presheaf $\underline{\Sigma}$ given by

$$\underline{S}_{\hat{P}} := \left\{ S_{\delta(\hat{P})_V} \subseteq \underline{\Sigma}_V \mid V \in \text{Ob}(\mathcal{V}(\mathcal{H})) \right\}. \tag{4}$$

This leads to the main achievement of the Döring-Isham approach, the map which "throws" the observable into a world of classical perspectives:

The *daseinisation* $\delta$ of projection operators $\hat{P} \in \mathcal{P}(\mathcal{H})$ is the mapping

$$\begin{aligned} \delta : \mathcal{P}(\mathcal{H}) &\to \text{Sub}_{\text{cl}}(\underline{\Sigma}) \\ \hat{P} &\mapsto \underline{S}_{\hat{P}}. \end{aligned} \tag{5}$$

The importance of the definition of daseinisation rests on the mapping between a projection, which in *quantum* physics is the representative of a proposition of the theory, and a sub-object of the 'state object' $\underline{\Sigma}$, the topos analogue of a subset of the state space, which is the *classical* notion of a proposition in physics:

$$\text{projection } \hat{P} \text{ ["a quantum mechanical statement"]} \xrightarrow{\delta}$$
$$\text{subobject } S_{\hat{P}} \text{ [topos analogue of a subspace or a "classical statement"]}$$

The above constructions determine the logic appropriate for quantum physics in topoi, for note that $\text{Sub}_{\text{cl}}(\underline{\Sigma})$ (the clopen subobjects of $\underline{\Sigma}$) is a *Heyting algebra*. The distributive law holds in all Heyting algebras:

$$x \wedge (y \vee z) \leftrightarrow (x \wedge y) \vee (x \wedge z) \tag{6}$$

Hence, it is valid in propositional quantum logic. The well-known laws below, however, do not hold:

$$x \vee \neg x \text{ and } \neg \neg x \to x. \tag{7}$$

The quantum logic of topos physics is *intuitionistic*.

## 1.2. Bohrification

An alternative, mathematically sophisticated version of the topos-theoretical approach is found in the work of Heunen, Landsman and Spitters [10, 11]. This alternative, known as 'Bohrification', utilizes the topos-theoretical generalisation of the notion of space, *locales* (cf. sec. 4 below). The quantum logic is then read off from the Heyting algebra structure of the open subsets of a locale $L$ (or, strictly, the frame $\mathcal{O}(L)$), identified as the state space of the system. We review the main characteristics of Bohrification in this subsection.

In the Döring-Isham approach, the context category was given by a family of commutative subalgebras $V$ of a non-commutative von Neumann algebra. The state object $\underline{\Sigma}$ was a functor in the topos, with $\underline{\Sigma}_V$ the Gelfand spectrum in the context *V*. The construction relied on the rich supply of projections available in von Neumann algebras. In the Bohrification approach, a family of commutative subalgebras of a non-commutative C*-algebra is used instead of the von Neumann algebras. These algebras are generally poor in projections, special cases (such as Rickart algebras or von Neumann algebras) excepted, so the former notion of a state is no longer useful. (The two approaches to topos physics, Döring-Isham and Bohrification, are compared in great detail in [17].)

Bohrification starts from the topos **Sets**$^{C(A)}$ of *covariant* functors, where $C(A)$ is the set of commutative C*-subalgebras of a C*-algebra *A*. The tautological functor $\underline{A} : C(A) \to$ **Sets**, which acts on objects as $\underline{A}(C) = C$, and on morphisms $C \subseteq D$ as the inclusion $\underline{A}(C) \to \underline{A}(D)$, is called the *Bohrification* of *A*.

Now consider the functor $\mathcal{T} :$ **CStar** $\to$ **Topos**, where **CStar** is the category of unital C*-algebras (with arrows defined as linear multiplicative functions which preserve the identity and the *-operation), and **Topos** is the category of topoi (with geometric morphisms as arrows). $\mathcal{T}$ is defined by $\mathcal{T}(A) =$ **Sets**$^{C(A)}$ on objects and $\mathcal{T}(f)^*(\underline{T})(D) = \underline{T}(f(D))$ on morphisms $f : A \to B$, with $\underline{T} \in$ **Sets**$^{C(B)}$ and $D \in C(A)$. ($\mathcal{T}(f)^*$ is the inverse image part of the geometric morphism $\mathcal{T}(f)$.) It can then be shown that $\underline{A}$ is a commutative C*-algebra in the topos $\mathcal{T}(A) =$ **Sets**$^{C(A)}$. This crucial result rests upon a general fact from topos theory:

**Fact.** *If* **Model**($\mathcal{T}$, **T**) *denotes the category of models of a geometric theory* $\mathcal{T}$ *in the topos* **T**, *there is an isomorphism of categories*

$$\mathbf{Model}\left(\mathcal{T},\ \mathbf{Sets}^{C(A)}\right) \approx \mathbf{Model}\left(\mathcal{T},\ \mathbf{Sets}\right)^{C(A)}. \tag{8}$$

This is a special case of lemma 3.13 in [11]. (For a proof, see cor. D1.2.14 in [13].)

The proof of the commutativity of $\underline{A}$ appeals to Kripke-Joyal semantics for Kripke topoi [11]. It also makes use of the axiom of dependent choice (**DC**), which holds in **Sets**$^{C(A)}$. Commutativity of $\underline{A}$ in **Sets**$^{C(A)}$ is proved by exploiting the proximity of the theory of C*-algebras to a *geometric* theory. In these theories, all statements have the form

$$\forall (\vec{x})[\psi(\vec{x}) \to \phi(\vec{x})]. \tag{9}$$

Here, $\psi$ and $\phi$ are positive formulae; i.e. formulae built by means of finite conjunctions and existential quantifiers. Thus, geometric theories are formulae with "finite verification" (see [14], ch. X for more about this notion). If the theory of abelian C*-algebras (Banach algebras with involution, and satisfying $\|a^*a\| = \|a\|^2$) had been a geometric theory, we could start from the following piece of information about $\underline{A}$:

$$\underline{A} \in \mathbf{Model}\left(\text{The theory of abelian } C*-\text{algebras},\ \mathbf{Sets}\right)^{C(A)}. \tag{10}$$

This is true by the definition of $\underline{A}$ as the tautological functor, and because $C(A)$ contains only commutative subalgebras. By the fact stated above, it would then seem follow that

$$\underline{A} \in \textbf{Model}\left(\text{The theory of abelian } C*-\text{algebras, } \textbf{Sets}^{C(A)}\right). \tag{11}$$

That is, $\underline{A}$ is an internal C*-algebra in the topos $\textbf{Sets}^{C(A)}$. However, the theory of abelian C*-algebras is not a geometric theory: the axiom of completeness (the convergence of any Cauchy sequence in the algebra) fails us. In order to circumvent this difficulty, the authors introduce the notion of a "pre-semi-C*-algebra". All C*-algebras are "pre-semi", and the theory of these algebras *is* geometric. Again, by appeal to the fact above, $\underline{A}$ is an internal abelian "pre-semi". It is then shown "by hand" that $\underline{A}$ is, in fact, an internal abelian C*-algebra.

Now recall that, in the topos **Sets**, there is an equivalence (the *Gelfand duality*) between the categories **cCStar** (the commutative C*-algebras) and **KHausTop** (the compact Hausdorff topological spaces). In turn, **KHausTop** is equivalent to the category **KRegLoc** of *compact regular locales* in **Sets**. Banaschewski and Mulvey [1] have shown that the equivalence **cCStar** ⇌ **KRegLoc** holds in any topos. We shall not give the details of this beautiful, but demanding construction, which recently has been improved by Coquand and Spitters.

Let $\underline{\Sigma}$ be the morphism from **cCStar** to **KRegLoc**$^{op}$ in $\textbf{Sets}^{C(A)}$. (The underlining, also of $\underline{\Sigma}$, is a reminder that objects and morphisms between them are now internal to this topos.) Consider the locale $\underline{\Sigma}(\underline{A})$, the *Gelfand spectrum* of $\underline{A}$ (which, as we noted, is commutative in $\textbf{Sets}^{C(A)}$). $\underline{\Sigma}(\underline{A})$ is the state space ion the Bohrification approach, corresponding to the state object $\underline{\Sigma}$ in the Isham-Döring model. Interestingly, the locale $\underline{\Sigma}(\underline{A})$ is pointfree for $A$ = **Hilb**($H, H$), with $H$ a Hilbert space of dimension greater than 2 ([11], theorem 4.10), and also for more general classes of C*-algebras. This is the Bohrified version of the Kochen-Specker theorem, which was formulated for topos physics by Isham and Butterfield.

The construction of $\underline{\Sigma}(\underline{A})$ is done by means of formal symbols for each self-adjoint element $a$ of $A$, but we shall not go into this. As usual for entities in topoi, the Gelfand spectrum $\underline{\Sigma}(\underline{A})$ may alternatively be given an *external* description, and it can be shown that $\underline{\Sigma}(\underline{A})$ is determined by the value taken at $\mathbb{C}$ (the algebra of complex numbers is the least member of $C(A)$). $\underline{\Sigma}(\underline{A})(\mathbb{C})$, denoted by $\Sigma_A$, is known as the *Bohrified state space* of $A$. We shall study a concrete example of an external state space when we apply topos methods within loop quantum gravity in sect. 4 below.

Finally, the frame (or Heyting algebra) $O(\Sigma_A)$ provides a new quantum logic, which may be compared with the old Birkhoff-von Neumann logic when the C*-algebra has enough projections. This is the case when $A$ is a Rickart C*-algebra (see [11], sec. 5, for definitions and results). The atomic propositions of the theory are identified with elements of $O(\Sigma_A)$, and the resulting logic is intuitionistic.

## 2. The C*-Algebra Formulation of Loop Quantum Gravity

As an application of the methods developed, we shall attempt to represent a particular version of quantum gravity, the theory of loop quantum gravity (LQG), within the topos-theoretical framework. LQG generalizes the canonical methods from standard quantum mechanics, so it seems to be a natural choice for a topos model.

The radical space-time structure of LQG is embedded in a standard differential manifold, and physics is implemented by representing the observables of the theory as operators on a Hilbert space. Therefore, the cumbersome name "modern canonical quantum general relativity" is also used. The real geometrical structure of the theory emerges because the operators of the theory are constructed in accordance with the principle of *diffeomorphism invariance*.

The operators corresponding to the geometric properties of the system are defined, and the attempt is made to deduce their spectral properties. In the case of the area and volume operators, the spectrum is claimed to be discrete, leading to the non-classical picture of a space composed of finite, indivisible "grains" or "quanta of gravity". The eigenvectors of the geometric operators form a basis of the corresponding Hilbert space $\mathcal{K}_{\text{Diff}}$, and the radical picture of space as a superposition of "spin networks" emerges, with the nodes representing volume grains and the links representing the adjacent areas. This is the real quantum geometry which underlies the arbitrary coordinatization of the manifold the theory starts from.

An excellent overview of LQG is available, Rovelli's *Quantum Gravity* [15], whereas technical details and proofs can be found in [16]. We shall not enter into the details of the standard formulation of LQG, because the topos-theoretical approach forces us to collect the operators of the theory in an algebra with the appropriate structure. For this, the construction will rely on the C*-algebra version of LQG introduced in [9], which we summarize below.

*Configuration space and state space.* Assume $\Sigma$ is a 3d submanifold of space-time $(M, g)$. We leave it open whether $\Sigma$ is differentiable, analytic or even, for some purposes, semi-analytic (see [16], p. 162, for an enumeration of demands on $\Sigma$). The reader may prefer to think of $\Sigma$ as the Cauchy surface on which we collect our physical data. The set of all *paths* in $\Sigma$ (equivalent up to endpoints, orientation-preserving reparametrizations and the deletion of retraced curves $c \circ c^{-1}$) is denoted by $\mathcal{P}$. We may regard $\mathcal{P}$ as a groupoid under composition of paths. (We shall only be able to compose paths when the second path takes off from the end point of the first, so the operation is only partial.) Informally, we say that *graphs v* are finite collections of independent edges, where an *edge* is a path with no crosses (but possibly closed). All paths are finite combinations of edges. A collection of edges is *independent* if the edges meet each other at most in the beginning and final points. Then $\mathcal{P}_v$ is the subgroupoid in $\mathcal{P}$ generated by the edges in the graph $v$.

**Definition 1** [Cf. [9]]   (i) Hom($\mathcal{P}$, $SO(3)$) *is the set of groupoid morphisms from the set of paths in $\Sigma$ into $SO(3)$.* (ii) Hom($\mathcal{P}_v$, $SO(3)$) *is the set of groupoid morphisms from the subgroupoid $\mathcal{P}_v$ to $SO(3)$. (Homomorphisms in this set will be denoted by $x_v$.)*

**Definition 2** [Cf. [9]]   (i) $\overline{\mathcal{A}} = $ Hom($\mathcal{P}$, $SO(3)$) *is the* set of generalized connections *(on a*

*principal SO*(3) *bundle with base manifold* $\Sigma$). (ii) $\overline{\mathcal{A}_v}$ = Hom($\mathcal{P}_v$, *SO*(3)).

The next step is to find a topology for the space $\overline{\mathcal{A}}$. Note that the sets $\overline{\mathcal{A}_v}$ may be identified with $SO(3)^{\#v}$ (where $\#v$ is the number of edges in $v$). $SO(3)$ is a compact Hausdorff space, so $\overline{\mathcal{A}_v}$ is compact Hausdorff too. Recall that the *Tychonov topology* on a direct product $X_\infty$ (of any cardinality) of topological spaces $X_v$ is the weakest topology such that the projections onto the component spaces are continuous. Also, by Tychonov's theorem, the direct product space of compact topological spaces is a compact topological space (in the Tychonov topology). Accordingly, the direct product space $\mathcal{A}_\infty := \prod_{v \subset \mathcal{P}} \overline{\mathcal{A}_v}$ is compact (it is also Hausdorff).

In order to situate $\overline{\mathcal{A}}$ within $\mathcal{A}_\infty$, we appeal to the notion of a projective limit. We let $\mathcal{L}$ be the set of subgroupoids of $\mathcal{P}$, and say that $\mathcal{P}_v \prec \mathcal{P}_{v'}$ (or simply $v \prec v'$) iff $\mathcal{P}_v$ is a subgroupoid of $\mathcal{P}_{v'}$. It can be shown that $\mathcal{L}$ is a partially ordered and directed set (the last point is not entirely trivial; see [16], th. 6.2.13). The set $\mathcal{L}$ is associated with a *projective family* $(\overline{\mathcal{A}_v}, p_{v'v})_{v \prec v' \in \mathcal{L}}$, where $p_{v'v} : \overline{\mathcal{A}_{v'}} (= \text{Hom}(\mathcal{P}_{v'}, SO(3))) \to \overline{\mathcal{A}_v} (= \text{Hom}(\mathcal{P}_v, SO(3)))$ is the projection of the groupoid $\overline{\mathcal{A}_{v'}}$ onto its subgroupoid $\overline{\mathcal{A}_v}$. The *projective limit* $\overline{\mathcal{A}}$ of the projective family $(\overline{\mathcal{A}_v}, p_{v'v})_{v \prec v' \in \mathcal{L}}$ is the subset of $\mathcal{A}_\infty$ given by

$$\overline{\mathcal{A}} = \{(x_v)_{v \in \mathcal{L}} \mid \forall \, v \prec v' \, (p_{v'v}(x_{v'}) = x_v)\}. \tag{12}$$

The elements $(x_v) \equiv (x_v)_{v \in \mathcal{L}}$ are called *nets*. Note that the sign $\overline{\mathcal{A}}$ here makes a second entrance. This is justified by the following lemma.

**Lemma 1** [cf. [16], th. 6.2.22]  *The set of generalized connections $\overline{\mathcal{A}}$ is the projective limit of the projective family* $(X_l, p_{l'l})_{l \prec l' \in \mathcal{L}}$.

We let $\pi_\gamma : \overline{\mathcal{A}} \to \overline{\mathcal{A}_v} \simeq SO(3)^{\#v}$ be the projection of a generalized connection $\overline{A}$ onto its family member in $\overline{\mathcal{A}_v}$ and stipulate

$$h_{\overline{A}}(\gamma) := h_\gamma(\overline{A}) := \pi_\gamma(\overline{A}). \tag{13}$$

We now define the topology on $\overline{\mathcal{A}}$ as the subspace topology of $\overline{\mathcal{A}}$ with respect to the Tychonov topology on $\mathcal{A}_\infty$. It can be shown (cf. Thiemann [16], th. 6.2.19) that $\overline{\mathcal{A}}$ is a closed subspace of $\mathcal{A}_\infty$, so $\overline{\mathcal{A}}$ is compact. $\overline{\mathcal{A}}$ is the configuration space upon which the states of the theory are to be defined. The configuration space is given the *Ashtekar-Lewandowski measure* $\mu_0$ (for a precise definition, see [16], def. 8.2.4). We then identify the state space $\mathcal{H}$ of the theory (as usual, we have the norm on $L_2(\overline{\mathcal{A}}, \mu_0)$ induced by the standard inner product):

**Definition 3** [Cf. [9]]  *The* state space $\mathcal{H}$ *is the Hilbert space* $L_2(\overline{\mathcal{A}}, \mu_0)$ *of measurable square-integrable functions over the space $\overline{\mathcal{A}}$ of generalized connections, where $\mu_0$ is the Ashtekar-Lewandowski measure.*

*The operators.* We now have the set of bounded operators on the state space, $B(L_2(\overline{\mathcal{A}}, \mu_0))$, at our disposal, and may proceed with the construction of the C*-algebra by picking the appropriate operators within this set. Firstly, we identify the set of configuration operators with the set of multiplicative operators corresponding to the continuous functions on the configuration space $\overline{\mathcal{A}}$:

**Definition 4** [Cf. [9]]   $T = \{T_f \in B(L_2(\overline{\mathcal{A}}, \mu_0)) \mid f \in C(\overline{\mathcal{A}})\}$ *is the* set of configuration operators *on the state space* $(L_2(\overline{\mathcal{A}}, \mu_0)$.

In order to define the flux or "momentum" operators of LQG, we need some notion of a surface in $\Sigma$. Following Fleischhack ([9], p. 22), we say that a subset $S$ of $\Sigma$ is a *quasi-surface* iff every edge $\gamma$ can be decomposed into a finite set of segments $\{\gamma_1, ..., \gamma_n\}$ such that the interior of any segment $\gamma_i$ is either included in $S$ or has no points in common with $S$. As we have not yet commited ourselves to a particular choice of smoothness properties for $\Sigma$ and the edges in $\Sigma$, this flexibility carries over into the definition of $S$. We shall, however, suppose that the surfaces have an orientation. For this purpose, we say that a quasi-surface $S$ is *oriented* if there exists a function $\sigma_S$ from the set of all parameterized paths to the set $\{-1, 0, 1\}$ such that

$$\sigma_S(\gamma) = \begin{cases} \pm 1 & \text{if } \gamma(0) \in S \text{ and } \gamma \text{ does not have an initial segment included in } S \\ 0 & \text{if } \gamma(0) \notin S \text{ or } \gamma \text{ has an initial segment included in } S. \end{cases} \qquad (14)$$

We also demand that, if $\gamma_1$ and $\gamma_2$ are paths such that $\gamma_1$ ends where $\gamma_2$ begins, then $\sigma_S(\gamma_1 \circ \gamma_2) = \sigma_S(\gamma_1)$ unless $\gamma_1$ ends on $S$. For a $\gamma$ which starts and ends on $S$ without crossing it, we demand $\sigma_S(\gamma) = \sigma_S(\gamma)$. This assures that $\sigma_S(\gamma_1) = \sigma_S(\gamma_2)$ for paths $\gamma_1$ and $\gamma_2$ which start on $S$ and have an initial segment in common. Another reasonable requirement is to set $\sigma_S(\gamma) = (-)^n \sigma_S(\gamma)$ for paths which start and end on $S$, after crossing it $n$ times. The function $\sigma_S$ is called the *intersection function*.

The quantum gravity momentum ought to be defined as a sort of generator of small translations in configuration space (the generalized connection space $\overline{\mathcal{A}}$). The next result is therefore important:

**Proposition 2** [[9], prop. 3.19]   *Given a quasi-surface $S$ and an intersection function $\sigma_S$, and let $\gamma$ be a path in $\Sigma$ which does not traverse the surface $S$. There is a unique map* $\Theta^{S,\sigma_S} : \overline{\mathcal{A}} \times \text{Maps}(\Sigma, SO(3)) \to \overline{\mathcal{A}}$ *such that*

$$h_{\Theta^{S,\sigma_S}(\overline{A},d)}(\gamma) = \begin{cases} d(\gamma(0))^{\sigma_S(\gamma)} h_{\overline{A}}(\gamma) d(\gamma(1))^{-\sigma_S(\gamma^{-1})} & \text{if the interior of } \gamma \text{ is not included in } S \\ h_{\overline{A}}(\gamma) & \text{if the interior of } \gamma \text{ is included in } S. \end{cases}$$

$\Theta^{S,\sigma_S}$ *is continuous if* $\text{Maps}(\Sigma, SO(3))$ *is given the product topology. We define* $\Theta_d^{S,\sigma_S} : \overline{\mathcal{A}} \to \overline{\mathcal{A}}$ *by*

$$\Theta_d^{S,\sigma_S}(\overline{A}) = \Theta^{S,\sigma_S}(\overline{A}, d).$$

*Then $\Theta_d^{S,\sigma_S}$ is a homeomorphism which preserves the measure $\mu_0$ on $\overline{\mathcal{A}}$.*

$\Theta_d^{S,\sigma_S}$ is the sought-for class of momentum operators. In most cases, $\Theta_d^{S,\sigma_S}$ will be unbounded, but this difficulty is quickly removed:

**Lemma 3** *For $(X, \mu)$ a compact Hausdorff space, $\mu$ a regular Borel measure on $X$ and $\psi : X \to X$ a continuous surjective map which leaves $\mu$ invariant, the pull-back map $\psi^* : C(X) \to C(X)$ can be extended to a unitary operator on $L_2(X, \mu)$.*

**Proof**  It suffices to prove that $\psi^*$ is an isometry of $L_2(X, \mu)$ onto $L_2(X, \mu)$. Note first that $\psi^*$ is an isometry on the linear subspace $C(X)$ of $L_2(X, \mu)$:

$$|\psi^* f|^2 = \int_X \overline{\psi^* f}\, \psi^* f\, d\mu = \int_X \overline{f}\, f \circ \psi\, d\mu = \int_X \overline{f}\, f\, d\mu = |f|^2.$$

By the Stone-Weierstrass theorem, $C(X)$ is dense in $L_2(X, \mu)$, so $\psi^*$ can be extended to $L_2(X, \mu)$. By continuity of $\psi$, $\psi^*$ is an isometry on $L_2(X, \mu)$. It is also onto, for given $f \in L_2(X, \mu)$, there is a sequence $\{f_n\}$ in $C(X)$ such that $f_n \to f$. But $\psi$ is surjective, so there is a convergent sequence $\{f_n'\}$ in $C(X)$ with $\psi^* f_n' = f_n' \circ \psi = f_n$ for each $n$. Let $f' = \lim f_n'$. Then $\psi^* f' = \psi^*(\lim f_n') = \lim \psi f_n' = \lim (f_n' \circ \psi) = \lim f_n = f$. □

**Lemma 4** *For $f \in C(X)$ and $\psi : X \to X$, the corresponding operators $T_f$ and $w \equiv \psi^*$ in $B(L_2(X, \mu))$ satisfy $w \circ T_f \circ w^{-1} = T_{w(f)}$.*

**Proof**  Assume $h \in C(X)$. Then $(T_{w(f)} \circ w)h = T_{w(f)}(w(h)) = T_{w(f)}(h \circ \psi) = w(f)(h \circ \psi) = (f \circ \psi)(h \circ \psi) = (f h) \circ \psi = w(f h) = w(T_f h) = (w \circ T_f)h$. But $C(X)$ is dense in $L_2(X, \mu)$, so the relation holds also for $h \in L_2(X, \mu)$. □

The configuration space $\overline{\mathcal{A}}$ with measure $\mu_0$ fulfills the conditions in lemma 3. Also, $\Theta_d^{S,\sigma_S}$ is a homeomorphism, hence surjective. Application of the proposition to the momentum operators $\Theta_d^{S,\sigma_S}$ now allows us to define the Weyl operators:

**Definition 5** [[9], def. 3.21] *Let $\Theta_d^{S,\sigma_S} : \overline{\mathcal{A}} \to \overline{\mathcal{A}}$ be a momentum operator as in prop. 2. The Weyl operator $w_d^{S,\sigma_S} : L_2(\overline{\mathcal{A}}, \mu_0) \to L_2(\overline{\mathcal{A}}, \mu_0)$ is defined as the pull-back of the momentum operator,*

$$w_d^{S,\sigma_S} := (\Theta_d^{S,\sigma_S})^*.$$

The next, central definition gives us the C*-algebra needed for the toposification of loop quantum gravity:

**Definition 6** [[9], def. 4.1]  *The* loop quantum gravity C*-algebra $\mathcal{W}$ *is the subalgebra of bounded operators in* $B(L_2(\overline{\mathcal{A}}, \mu_0))$ *generated by* $T = \{T_f \in B(L_2(\overline{\mathcal{A}}, \mu_0)) \mid f \in C(\overline{\mathcal{A}})\}$ *(where $T_f$ is the multiplicative operator associated with f) and the Weyl operators $w_d^{S,\sigma_S}$.*

We say that $T$ is the *set of position* (or *configuration*) *operators*. The choice of $C(\overline{\mathcal{A}})$ for this purpose is analogous to the definition of position operators in quantum mechanics (cf. [9], p. 15 and [15], p. 199). $\mathcal{W}$, as defined above, fulfills the demands on a (concrete) C*-algebra. (The norm of the algebra is simply the operator norm in $B(L_2(\overline{\mathcal{A}}, \mu_0)$, which satisfies the additional norm condition $\|A^*A\| = \|A\|^2$.)

## 3. The Bohrification of LQG

We now switch from the familiar topos **Sets** and proceed with the investigation in the less explored surroundings of $\textbf{Sets}^{C(\mathcal{W})}$. Firstly, note that we shall need the following supplement to the Bohrification method of subsection 1.2. According to Bohr's thesis, observation is always filtered through classical concepts. However, one may argue that it is not the observables themselves that are of primary importance, but rather their evolution. Of main interest in particle physics is the calculation of transition probabilities when one or several particles approach an interaction region from infinitely far off, and leave again at infinity. In fact, these probability distributions (the cross-sections) are all that is measured. For example, suppose that $\Psi_{in}$ is the incoming particle state, and we want to find the probability that the outgoing state is $\Psi_{out}$. The amplitude for development from $\Psi_{in}$ to $\Psi_{out}$ will then be given by the quantity $\langle \Psi_{out} \mid e^{-iHt} \Psi_{in} \rangle$, where $U \equiv e^{-iHt}$ is the unitary Weyl operator corresponding to the Hamiltonian $H$.

So the contexts of our topos model ought to be subalgebras of a Weyl algebra $A_W$ (called $\mathcal{W}$ in the LQG case in sec. 2). The functor $\underline{A_W}$, the counterpart of $A_W$ in the topos, may then be called the 'Weylification' of the original, untamed algebra $A$ generated by the "position" and "momentum" operators of $A$. Quite apart from the present topic, quantum gravity, it would be of interest to see to if Weylification modifies the constructions of the topos-theoretical approach to quantum physics.

We shall now apply topos-theoretical methods to loop quantum gravity, as represented by Fleischhack's non-commutative C*-algebra $\mathcal{W}$. (See ch. 3 of [3] for more details on the topos model of LQG.) Relying on the Bohrification method sketched in sec. 2, our first step is the construction of the *commutative* algebra $\underline{\mathcal{W}}$ in a certain topos, namely

**Definition 7**  *Let $C(\mathcal{W})$ be the partially ordered set of commutative C*-subalgebras of $\mathcal{W}$. Then $\tau_{\mathcal{W}} := \textbf{Sets}^{C(\mathcal{W})}$ (or $[C(\mathcal{W}), \textbf{Sets}]$) is the topos of covariant functors from the category $C(\mathcal{W})$ to the category* **Sets**.

Note that the category structure $C(\mathcal{W})$ stems from the partial order on $C(\mathcal{W})$ given by inclusion: there is a morphism $C \to D$ iff $C \subset D$. It is clear that $C(\mathcal{W})$ contains non-trivial commutative subalgebras of $\mathcal{W}$. Below, we mention a few.

**Example 1** Let $W_T$ be the subalgebra of $\mathcal{W}$ generated by the set of configuration operators $T = \{T_f \in B(L_2(\overline{\mathcal{A}}, \mu_0)) \mid f \in C(\overline{\mathcal{A}})\}$. Then $W_T$ is a commutative C*-subalgebra of $\mathcal{W}$.

**Example 2** [Cf. [9], cor. 3.23] Let $(S, \sigma_S)$ be an oriented quasi-surface, and let $D$ be a set of functions $d : \Sigma \to SO(3)$ such that $d_1 d_2 = d_2 d_1$ for all $d_1, d_2 \in D$. Define $W_{S,D}$ as the subalgebra of $\mathcal{W}$ generated by the set of all operators $w_d^{S,\sigma_S}$ with $d \in D$. Then $w_{d_1}^{S,\sigma_S} w_{d_2}^{S,\sigma_S} = w_{d_1 d_2}^{S,\sigma_S} = w_{d_2}^{S,\sigma_S} w_{d_1}^{S,\sigma_S}$. Indeed, assume $f \in L_2(\overline{\mathcal{A}}, \mu_0)$ and let $\{f_n\}$ be a sequence in $C(X)$ such that $f_n \to f$. We have, for each $n$,

$$(w_{d_1}^{S,\sigma_S} w_{d_2}^{S,\sigma_S}) f_n = w_{d_1}^{S,\sigma_S}(f_n \circ \Theta_{d_2}^{S,\sigma_S}) = (f_n \circ \Theta_{d_2}^{S,\sigma_S}) \circ \Theta_{d_1}^{S,\sigma_S} = f_n \circ \Theta_{d_1 d_2}^{S,\sigma_S} = w_{d_1 d_2}^{S,\sigma_S} f_n.$$

By taking the limit, we find that these operators are commutative over all of $L_2(\overline{\mathcal{A}}, \mu_0)$. Above, we used the relation

$$\Theta_{d_2}^{S,\sigma_S} \circ \Theta_{d_1}^{S,\sigma_S} = \Theta_{d_1 d_2}^{S,\sigma_S}.$$

This can easily be derived from the definition (in prop. 2) of $\Theta$. Consider e.g. the case where a path $\gamma$ leaves the surface $S$ in the positive direction without return. Then

$$h_{\Theta_{d_1 d_2}^{S,\sigma_S}(\overline{A})}(\gamma) = d_1(\gamma(0)) d_2(\gamma(0)) \, h_{\overline{A}}(\gamma) = d_1(\gamma(0)) \, h_{\Theta_{d_2}^{S,\sigma_S}(\overline{A})}(\gamma) = h_{\Theta_{d_1}^{S,\sigma_S}\left(\Theta_{d_2}^{S,\sigma_S}(\overline{A})\right)}(\gamma) = h_{\Theta_{d_1}^{S,\sigma_S} \circ \Theta_{d_2}^{S,\sigma_S}(\overline{A})}(\gamma).$$

This shows that $W_{S,D}$-algebras are commutative subalgebras of $\mathcal{W}$. The proof depended crucially on the commutativity of the "translator functions" $d$. For a given surface $S$, the algebra $W_S$ generated by the set $\bigcup_{D \text{ comm}} W_{S,D}$ will in general not be commutative.

**Example 3** The algebras $W_T$ and $W_{S,D}$ belong to the configuration operator and momentum operator region, respectively. We might wonder if there are commutative subalgebras of $\mathcal{W}$ which combine these regions. Let $S$ be a given quasi-surface. If we apply lemma 4 to the operators $w_d^{S,\sigma_S}$ and $T_f$, we see that the relation $w_d^{S,\sigma_S} \circ T_f \circ \left(w_d^{S,\sigma_S}\right)^{-1} = T_{w_d^{S,\sigma_S}(f)}$ holds in $\mathcal{W}$. Therefore, whenever $f = w_d^{S,\sigma_S}(f)$, this reduces to a commutative relation

$$w_d^{S,\sigma_S} \circ T_f = T_f \circ w_d^{S,\sigma_S}. \tag{15}$$

Writing $f \equiv T_f$ for the multiplicative operator and calculating with $h \in L_2(\overline{\mathcal{A}}, \mu_0)$, this amounts to the demand that

$$w_d^{S,\sigma_S}(fh) = f \cdot w_d^{S,\sigma_S}(h). \tag{16}$$

Note that the operators in $\mathcal{W}$ may also have commutative relations with operators *outside* the algebra. Thus, lemma 5 below gives the commutative instances of Fleischhack's constructions of "graphomorphisms". Assume that $\phi : \Sigma \to \Sigma$ is a bijective function such that $\phi(S) = S$ and that $\phi$ does not switch the orientation of the surface $S$. The smoothness properties of $\phi$ should correspond to those of the paths (which we have left undecided). Now $\phi$ induces a map $\phi_\mathcal{P} : \mathcal{P} \to \mathcal{P}$ on $\mathcal{P}$, namely $\phi_\mathcal{P}(\gamma) = \phi \circ \gamma$. Also, we may define still another map $\phi_{\overline{\mathcal{A}}}$, this time on the connections in $\overline{\mathcal{A}}$:

$$\phi_{\overline{\mathcal{A}}}(\overline{A})(\gamma) := h_{\overline{A}}\!\left(\phi^{-1} \circ \gamma\right). \tag{17}$$

It can been shown (by a proof similar to [9], prop. 3.31) that $\phi_{\overline{\mathcal{A}}}$ is a homeomorphism. The final step is to define the "external" operator $\alpha_\phi : C(\overline{\mathcal{A}}) \to C(\overline{\mathcal{A}})$ by

$$\alpha_\phi(f) := f \circ \phi_{\overline{\mathcal{A}}}^{-1}. \tag{18}$$

Again, the domain of $\alpha_\phi$ may be extended to all of $L_2(\overline{\mathcal{A}}, \mu_0)$ by lemma 3.

**Lemma 5** [Cf. [9], prop. 3.34]  *For a function $d : \Sigma \to SO(3)$ such that $d = d \circ \phi^{-1}$, we have the commutative relation*

$$w_d^{S,\sigma_S} \circ \alpha_\phi = \alpha_\phi \circ w_d^{S,\sigma_S}. \tag{19}$$

**Proof**  For $f \in C(\overline{\mathcal{A}})$, $\overline{A} \in \overline{\mathcal{A}}$, we have

$$\{[w_d^{S,\sigma_S} \circ \alpha_\phi](f)\}(\overline{A}) = \{w_d^{S,\sigma_S}(f \circ \phi_{\overline{\mathcal{A}}}^{-1})\}(\overline{A}) = [f \circ \phi_{\overline{\mathcal{A}}}^{-1} \circ \Theta_d^{S,\sigma_S}](\overline{A}) \quad (*)$$

and

$$\{[\alpha_\phi \circ w_d^{S,\sigma_S}](f)\}(\overline{A}) = \{\alpha_\phi(f \circ \Theta_d^{S,\sigma_S})\}(\overline{A}) = [f \circ \Theta_d^{S,\sigma_S} \circ \phi_{\overline{\mathcal{A}}}^{-1}](\overline{A}). \quad (**)$$

There are now several instances to consider. As in example 2, we give the proof for the case where a path $\gamma$ leaves the surface $S$ in the positive direction without return. Then, by our assumption on $d$,

$$h_{\Theta_d^{S,\sigma_S}(\overline{A})}(\gamma) = d(\gamma(0))\, h_{\overline{A}}(\gamma) = (d \circ \phi^{-1})(\gamma(0))\, h_{\overline{A}}(\gamma) = d(\phi^{-1}(\gamma(0)))\, h_{\overline{A}}(\gamma) = d(\phi^{-1}(\gamma(0))) \cdot h_{\phi_{\overline{\mathcal{A}}}^{-1}(\overline{A})}(\phi^{-1} \circ \gamma).$$

The last step follows because $\phi$ maps $S$ to $S$ and because, by (17),

$$h_{\phi_{\overline{\mathcal{A}}}^{-1}(\overline{A})}(\phi^{-1} \circ \gamma) = \phi_{\overline{\mathcal{A}}}\!\left(\phi_{\overline{\mathcal{A}}}^{-1}(\overline{A})\right)(\gamma) = h_{\overline{A}}(\gamma).$$

From prop. 2 we then have

$$h_{\Theta_d^{S,\sigma_S}(\overline{A})}(\gamma) = h_{\Theta_d^{S,\sigma_S}\left[\phi_{\overline{\mathcal{A}}}^{-1}(\overline{A})\right]}(\phi^{-1} \circ \gamma).$$

But, by the definition of the map $\phi_{\overline{\mathcal{A}}}$,

$$h_{\overline{A}}(\phi^{-1} \circ \gamma) = h_{\phi_{\overline{\mathcal{A}}}(\overline{A})}(\gamma).$$

Putting the last two steps together, we have

$$h_{\Theta_d^{S,\sigma_S}(\overline{A})}(\gamma) = h_{\phi_{\overline{\mathcal{A}}}(\overline{A})\left[\Theta_d^{S,\sigma_S}\left[\phi_{\mathcal{A}}^{-1}(\overline{A})\right]\right]}(\gamma).$$

This shows, finally, that

$$\Theta_d^{S,\sigma_S} = \phi_{\overline{\mathcal{A}}}(\overline{A}) \circ \Theta_d^{S,\sigma_S} \circ \phi_{\mathcal{A}}^{-1}.$$

The remaining cases for paths $\gamma$ are similar. By (*) and (**) above, this suffices to prove commutativity of $w_d^{S,\sigma_S}$ and $\alpha_\phi$ for $f \in C(\overline{\mathcal{A}})$. The general result then follows by the density of $C(\overline{\mathcal{A}})$ in $L_2(\overline{\mathcal{A}}, \mu_0)$. □

Ought we to include operators like the $\alpha_\phi$'s in the algebra $\mathcal{W}$? We will return to this question in section 5 below.

Finally, we have

**Definition 8** $\underline{\mathcal{W}}$ *is the tauological functor* $\underline{\mathcal{W}} : C(\mathcal{W}) \to$ **Sets** *such that* $C \mapsto C$, *and* $C \subset_{C(\mathcal{W})} D \mapsto C \subset_{\text{Sets}} D$ *for morphisms.*

As a special case of the result proven in [10], it holds that $\underline{\mathcal{W}}$ is a commutative C*-algebra in the topos $[C(\mathcal{W}), \text{Sets}]$. The same authors then apply the constructive Gelfand duality of Banaschewski and Mulvey in order to find the Gelfand spectrum $\underline{\Sigma}(\underline{A})$ of a commutative algebra $\underline{A}$ in the topos $[C(A), \text{Sets}]$ (cf. sec. 2). The computation of this spectrum has been greatly clarified in the general case in [17]. We will seek out its consequences for the Gelfand spectrum $\underline{\Sigma}(\underline{\mathcal{W}})$ (hereafter denoted by $\underline{\Sigma}$) of the LQG algebra $\underline{\mathcal{W}}$ in $[C(\mathcal{W}), \text{Sets}]$. Our aim is to deduce the sobriety properties of the external description of this functor (sec. 4 below).

## 4. Topological Properties of the State Space

In order to prove results about the sobriety of the Gelfand spectrum, it will be advantageous to rely on the *external* description of $\underline{\Sigma}(\underline{\mathcal{W}})$.

**Definition 9** [cf. [17]] *The* external Gelfand spectrum $\Sigma$ *of* $\mathcal{W}$ *is the set* $\{(C, \lambda) \mid C \in C(\mathcal{W}), \lambda \in \Sigma_C$ *(the Gelfand spectrum of the commutative subalgebra C)} with topology* $O\Sigma$ *such that*
$U \in O\Sigma$ *iff (1)* $U_C \equiv \{\lambda \in \Sigma_C \mid (C, \lambda) \in U\}$ *is open (in the weak\*-topology of $\Sigma_C$), and (2) if $\lambda \in U_C$, $C \subseteq C'$ and $\lambda'|_C = \lambda$ for $\lambda' \in \Sigma_{C'}$, then $\lambda' \in U_{C'}$.*

As an easy consequence of the definition, we may characterize the closed sets:

**Lemma 6** *A set V is closed in $\Sigma$ iff (1) $V_C$ is closed (in the weak*-topology of $\Sigma_C$) for all $C \in C(\mathcal{W})$ and (2) if $\lambda \in V_C$ and $D \subseteq C$ then $\lambda|_D \in V_D$.*

**Proof** Assume first that $V$ is closed in $\Sigma$. Then $\Sigma \setminus V$ is open, so $(\Sigma \setminus V)_C = \Sigma_C \setminus V_C$ is open in $\Sigma_C$. Hence, $V_C$ is closed in $\Sigma_C$ for all $C$. Let $\lambda \in V_C$ and $D \subseteq C$, and assume that $\lambda|_D \notin V_D$. Then $\lambda|_D \in (\Sigma \setminus V)_D$, which is open in $\Sigma_D$. But $\Sigma \setminus V$ is open in $\Sigma$, so $\lambda \in V_C \subseteq \Sigma_C$ implies that $\lambda \in (\Sigma \setminus V)_C = \Sigma_C \setminus V_C$, which contradicts $\lambda \in V_C$. So closedness implies condition (2) also. Implication in the opposite direction may be proven in a similar manner. □

Now, $\Sigma$ is the external description of the Gelfand spectrum $\underline{\Sigma}(\underline{\mathcal{W}})$:

**Proposition 7** [cf. [17], cor. 2.18] *The projection $\pi : \Sigma \to C(\mathcal{W})$ given by $\pi(C, \lambda) = C$ is isomorphic to $\underline{\Sigma}(\underline{\mathcal{W}})$ as a locale.*

The proof was completed recently by Wolters [17], and we shall not repeat it here. We should, however, use this opportunity to clarify a few points with respect to the internal/external distinction in topoi. Recall (cf. [14], ch. IX) that a *locale* is an object of the category **Locales**, the opposite of the category of frames, **Frames**. A *frame* is a lattice with all finite meets and all joins which satisfies the infinite distribution law

$$U \wedge \left( \bigvee_i V_i \right) = \bigvee_i (U \wedge V_i). \tag{20}$$

If $X$ is a locale, one usually denotes the corresponding frame by $\mathcal{O}(X)$. A map $f : X \to Y$ between locales corresponds to a frame map denoted by $f^{-1} : \mathcal{O}(Y) \to \mathcal{O}(X)$. A *point* $p^*$ in a frame $F$ is a map $p^* : F \to \{0, 1\} = \mathcal{O}(*)$. (Hence, for $F = \mathcal{O}(X)$, the open sets of a space $X$, $p \in X$ defines a point $p^*$ in $f$ if we set

$$p^*(U) = 1 \text{ iff } p \in U.$$

We also say that $\mathrm{Pt}(F)$ are the points in $F$ with open sets $\mathrm{Pt}(U) \equiv \{p^* \mid p^*(U) = 1\}$. A frame $F$ is *spatial* if it is isomorphic (as a frame) to $\mathcal{O}(\mathrm{Pt}(F))$. Dually, a topological space $X$ is *sober* if it is homeomorphic to $\mathrm{Pt}(\mathcal{O}(X))$.

An *internal frame* (or a *frame object*) of a topos is an object $F$ in the topos together with arrows $\wedge : F \times F \to F$ and $\vee : F \times F \to F$ such that the usual lattice identities and the distribution law (20) can be translated into commutative diagrams. In this sense, $\underline{\Sigma}(\underline{\mathcal{W}})$ is an internal locale of the topos $[C(\mathcal{W}), \mathbf{Sets}]$. Now note that the projection $\pi : \Sigma \to C(\mathcal{W})$ is a map between locales; that is, $\pi^{-1} : \mathcal{O}C(\mathcal{W}) \to \mathcal{O}\Sigma$ is a map between frames. Thus, it is claimed in prop. 7 that $\pi^{-1}$ is a *frame isomorphism*.

*Relationship with the Döring-Isham formalism.* The external spectrum $\Sigma$ looks quite similar to the state object $\underline{\Sigma}$ as defined in the Döring-Isham approach referred to in sec. 1. (This presupposes that we replace the von Neumann algebra $\mathcal{V}(\mathcal{H})$ with a C*-algebra $A$, so the daseinisation procedure is no longer available.) Indeed, the connection can be made precise by the following piece of category theory:

**Lemma 8** $\underline{\Sigma}$, *regarded as a category with morphisms* $(C, \lambda) \to (C', \lambda|_{C'})$ *for* $C' \subseteq C$, *is the category of elements of the state object* $\underline{\Sigma}$; *briefly*,
$$\Sigma = \int_{C(\mathcal{W})} \underline{\Sigma}.$$

**Proof** This follows directly from the definition of a category of elements ([14], p. 41). According to the definition, the objects of $\int_{C(\mathcal{W})} \underline{\Sigma}$ are all pairs $(C, \lambda)$ with $C \in A$ and $\lambda \in \underline{\Sigma}(C) = \Sigma_C$, the Gelfand spectrum of $C$, and the morphisms are $(C, \lambda) \to (C', \underline{\Sigma}(C' \to C)(\lambda)) = (C', \lambda|_{C'})$ for $C' \subseteq C$. □

The sobriety of a topological space may also be characterized in the following manner: we shall say that a topological space $S$ is *sober* iff every nonempty irreducible closed subset $V \subseteq S$ is the closure of a unique point $s$; explicitly, $V = \overline{\{s\}}$ ([14], p. 477); recall that a closed set is *irreducible* if it is not the union of two smaller closed subsets).

In the next theorem, the 3d *surface* $\Sigma$ should not be confused with the *Gelfand spectrum* $\Sigma$. We also assume, as before, that surfaces and paths have matching smoothness properties. For the following theorem, we choose one among several reasonable ways of simplying the structure of $\Sigma$.

**Theorem 9** *For* $\Sigma \simeq \mathbb{R} \times S$, *the external Gelfand spectrum* $\Sigma \equiv \Sigma(\mathcal{W})$ *is not sober*.

**Proof** First note that if $(C, \lambda)$ is any point in $\Sigma$, its closure is

$$\overline{\{(C, \lambda)\}} = \{(D, \lambda') \mid D \subseteq C \wedge \lambda' = \lambda|_D\}. \quad (*)$$

Indeed, if we write $X = \{(D, \lambda') \mid D \subseteq C \wedge \lambda' = \lambda|_D\}$, the singleton set $X_D = X \cap \Sigma_D = \{(D, \lambda')\}$ is trivially closed in $\Sigma_D$ (under the weak*-topology), and, if $\lambda' \in X_D$ and $E \subseteq D$, then $\lambda' = \lambda|_D$, so $\lambda'|_E = (\lambda|_D)|_E = \lambda|_E \in X_E$, so $X$ is closed by lemma 6. But if $Y$ is an arbitrary closed set which contains $(C, \lambda)$, by closedness $Y$ contains all $(D, \lambda')$ such that $D \subseteq C$ and $\lambda' = \lambda|_D$, so $X \subseteq Y$. That is, $X$ is the closure of $\{(C, \lambda)\}$.

We shall construct an irreducible closed subset $X^*$ of $\Sigma(\mathcal{W})$ which is not of the form (*), thereby proving that $\Sigma(\mathcal{W})$ is not sober. By Wolters ([17], lemma 2.24), irreducibility of a closed subset $V$ of $\Sigma(\mathcal{W})$ is equivalent to the following conditions:

1. For all $C \in C(\mathcal{W})$, $V_C \equiv V \cap \Sigma_C$ is either empty or singleton, and

2. For all nonempty $V_C$ and $V_{C'}$, there exists $C''$ such that $C, C' \subseteq C''$ and $V_{C''}$ is nonempty.

We shall now simplify the structure of $\Sigma$ by assuming that it has the topology $\Sigma = \mathbb{R} \times S$ for an arbitrary 2d manifold $S$. We may then use a sequence $\{S_i\}_{i \in \mathbb{N}}$ of non-intersecting surfaces in $\Sigma$ as the basis for our construction of $X^*$.

For each $i$, we now pick a Weyl operator $w_d^{S_i, \sigma_{S_i}}$ (with $d \in \text{Maps}(\Sigma, SO(3))$ an arbitrary mapping), and define a sequence $\{V_i\}$ of subalgebras of $\mathcal{W}$:

$$V_i = \text{the C*-algebra generated by } \{1, w_d^{S_1, \sigma_{S_1}}, \ldots, w_d^{S_i, \sigma_{S_i}}\}.$$

For a given $i$, assume that $w_d^{S_j, \sigma_{S_j}}, w_d^{S_k, \sigma_{S_k}} \in V_i$. Then

$$w_d^{S_j, \sigma_{S_j}} w_d^{S_k, \sigma_{S_k}} = w_d^{S_k, \sigma_{S_k}} w_d^{S_j, \sigma_{S_j}}.$$

(This is lemma 3.26 in [9].) Thus $V_i$ is commutative, so $V_i \in C(\mathcal{W})$ for each $i$.

Now assume that a character $\lambda_n$ has been constructed on $V_n$ for a given $n$ (that is, a multiplicative linear map $\lambda_n : V_n \to \mathbb{C}$). We may then extend $\lambda_n$ to a character on $V_{n+1}$ by defining $\lambda_{n+1}(w_d^{S_{n+1}, \sigma_{S_{n+1}}}) = c$ for some complex number $c$. (As multiplicative linear maps on a C*-algebra have norm 1 and the Weyl operators are unitary, we must demand that $c$ is on the unit circle.) Consider the set

$$X^* = \{(C, \lambda) \mid \text{there is some } n \in \mathbb{N} \text{ such that } V_n \subseteq C \subseteq V_{n+1} \wedge \lambda = \lambda_{n+1}|_C\}.$$

It follows that $X_C^*$ is either a singleton or the empty set, hence closed in $\Sigma_C$. Also, if $X_C^*$ is nonempty then $X_C^* = \{(C, \lambda)\}$, so $\lambda = \lambda_{n+1}|_C$ for some $n$. But then $D \subseteq C$ implies $\lambda|_D = (\lambda_{n+1}|_C)|_D = \lambda_{n+1}|_D \in X_D^*$. By lemma 6 again, $X^*$ is closed. Irreducibility of $X^*$ is likewise an easy consequence. We just noted that condition 1 holds, and for nonempty $X_C^*$ and $X_{C'}^*$ there are $n, n'$ such that $C \subseteq V_n$ and $C' \subseteq V_{n'}$. If we let $n^* = \max\{n, n'\}$, we have $C, C' \subseteq V_{n^*}$, and $V_{n^*}$ contains $\lambda_{n^*}$, hence is nonempty. Condition 2 above is thus satisfied.

Finally, we see that $X^*$ is not the closure of a unique point in $\Sigma(\mathcal{W})$. Assume, to the contrary, that $(F, \lambda|_F)$ is a point such that $X^* = \overline{\{(F, \lambda|_F)\}} = \{(D, \lambda') \mid D \subseteq C \wedge \lambda' = (\lambda|_F)|_D\}$. Note that $\bigcup_i V_i$ is a commutative subalgebra of $\mathcal{W}$, with a character $\lambda_\cup$ given by stipulating that $\lambda_\cup(w_d^{S_i, \sigma_{S_i}}) = \lambda_i(w_d^{S_i, \sigma_{S_i}})$ for each $i$. By definition of $X^*$, $X^*_{\bigcup V_i} = X^* \cap \Sigma_{\bigcup V_i}$ $= \emptyset$, so $(\bigcup_i V_i, \lambda_\cup) \notin X^*$. From the assumption that $X^* = \overline{\{(F, \lambda)\}}$ we also know that there is no $n$ such that $V_n \subseteq F \subseteq V_{n+1}$. Yet $(V_i, \lambda_i) \in \overline{\{(F, \lambda)\}}$ for all $i$, which implies that $V_i \subset F$ and $\lambda_i = \lambda_\cup|_{V_i}$ for all $i$. Hence, $\bigcup_i V_i \subseteq F$ and $\lambda_\cup = \lambda|_F|_{\bigcup_i V_i}$, from which it immediately follows that $(\bigcup_i V_i, \lambda_\cup) \in \overline{\{(F, \lambda|_F)\}}$.

We have a contradiction, so the irreducible closed subset $X^*$ is not the closure of a unique point, and, accordingly, $\Sigma$ is not a sober space. □

We say that $C(\mathcal{W})$ satisfies the *ascending chain condition* iff for every chain $C_1 \subseteq C_2 \subseteq C_3 \subseteq ...$ of contexts in $C(\mathcal{W})$, there is an $n$ such that $C_{m+1} = C_m$ for all $m \geq n$ ([17], th. 2.25).

**Corollary 10** *The algebra $\mathcal{W}$ does not satisfy the ascending chain condition.*

**Proof** Immediate from the construction in the proof of th. 9, or by noting that soberness is a consequence of the ascending chain condition ([17], th. 2.25). □

Theorem 9, then, brings together concepts from the arenas of loop quantum gravity and topos physics. Let us comment briefly on the status of this result. As a property of topological spaces, sobriety is situated between $T_0$ (the Kolmogorov condition) and $T_2$ (the Hausdorff condition) ([14], p. 477). Intuitively, the soberness of a space implies that if we continue to split a closed set into closed proper subsets, the process will only terminate at sets which are the closures of singleton sets. The $L_2$ state spaces familiar from quantum mechanics are Hausdorff spaces, and therefore sober. The non-sober state space $\Sigma$ above may be seen as a generalized (pointfree) space by noting that the functors $X \mapsto O(X)$ and $Pt(F) \hookleftarrow F$ give an equivalence between the categories (cf. [12])

$$\textbf{Sober spaces} \simeq \textbf{Spatial frames}^{op}.$$

Above we defined the category of pointfree spaces, **Locales**, as the opposite of the category of frames:

$$\textbf{Locales} := \textbf{Frames}^{op}.$$

From [14] (p. 480f) it now follows that, even if the external Gelfand spectrum $\Sigma$ is non-sober, the associated frame $O(\Sigma)$ is spatial, or, differently phrased, the locale has "enough points". Thus, the scarcity of points (non-sobriety) in the external space does not rule out the availability of corresponding spatial frames (locales) in the topos, where the notion of a locale emerges as a proxy for the notion of a space. The duality of external non-sobriety and internal sobriety seems to be a topological counterpart to the algebraic duality between the non-commutative algebra $\mathcal{W}$ in **Sets** and the commutativity of its representative $\underline{\mathcal{W}}$ in the topos $\tau_{\mathcal{W}}$. It should be explored further.

# 5. The Requirements of Diffeomorphism and Gauge Invariance

As a quantised theory of general relativity, LQG should fulfill the requirements of gauge invariance (under the Poincaré group for the full-blown theory) and diffemomorphism invariance (full freedom of coordinate choice). We must now ask how these invariance types are to be understood within topos physics. Focusing on the diffeomorphism case, the following interpretation is suggested. Note first that we may associate any diffeomorphism $\phi : \Sigma \to \Sigma$ with the *-morphism $A_\phi : \mathcal{W} \to \mathcal{W}$ defined by the following action on all generators $T_f$ and $w_d^{\sigma_S}$ of $\mathcal{W}$:

$$\left(A_\phi(T_f)\right)(\overline{A}) = T_f\left(\phi(\overline{A})\right) \quad \left(\text{for } \overline{A} \in \overline{\mathcal{A}}\right), \tag{21}$$

$$A_\phi\left(w_d^{S,\sigma_S}\right) = w_{\phi(d)}^{\phi(S),\phi(\sigma_S)}. \tag{22}$$

Note that here, as in lemma 5, we use the lifting of $\phi$ to a map $\phi \equiv \phi_{\overline{\mathcal{A}}} : \overline{\mathcal{A}} \to \overline{\mathcal{A}}$ by stipulating that $\phi(\overline{A})(\gamma) \equiv \pi_{\phi^{-1}\circ\gamma}(\overline{A})$, with $\pi_{\phi^{-1}\circ\gamma}$ the projection onto the component space $\overline{\mathcal{A}_{\phi^{-1}\circ\gamma}}$ of the path $\phi^{-1} \circ \gamma$ (that is, $\phi_{\overline{\mathcal{A}}}(\overline{A})(\gamma) = h_{\overline{A}}(\phi^{-1} \circ \gamma)$). We also write $\phi(S) \equiv \phi \circ S$; $\phi(d) = d \circ \phi^{-1}$; and $\phi(\sigma_S)(\gamma) = \sigma_{\phi^{-1}(S)}(\phi^{-1} \circ \gamma)$ (cf. [9], def. 3.25).

Fleischhack [9] shows that we can associate the diffeomorphisms $\phi : \Sigma \to \Sigma$ with operators $\alpha_\phi$ on $L_2(\overline{\mathcal{A}}, \mu_0)$ in a natural manner (cf. sec. 3 above): for any $f \in C(\overline{\mathcal{A}})$, we let $\alpha_\phi : C(\overline{\mathcal{A}}) \to C(\overline{\mathcal{A}})$ be given by $\alpha_\phi(f) \equiv f \circ \phi_{\overline{\mathcal{A}}}^{-1} \equiv f \circ \phi^{-1}$, the pullback of $\phi^{-1}$ (the lifting of $\phi^{-1}$ to $\overline{\mathcal{A}}$). Applying lemma 3, we extend $\alpha_\phi$ to a unitary operator on $L_2(\overline{\mathcal{A}}, \mu_0)$. As the operators $\alpha_\phi$ merely reflect a switch of coordinates and have no observational content, we have chosen not to include them in the Weyl algebra $\mathcal{W}$.

For a subalgebra $C$, we denote by $\phi C$ the algebra generated by the image $A_\phi(C)$.

**Definition 10** *A context (that is, a commutative subalgebra) $C \in C(\mathcal{W})$ is diffeomorphism invariant if, for any diffeomorphism $\phi : \Sigma \to \Sigma$, the algebra $\phi C$ is a commutative subalgebra in $C(\mathcal{W})$. If this holds for all contexts $C$, we say that $C(\mathcal{W})$ itself is diffeomorphism invariant.*

Keep in mind that, just as when we were discussing paths and surfaces, we do not want to commit ourselves to a particular choice of diffeomorphism type. (Fleischhack [9] considers the "stratified analytic diffeomorphisms".) The definition is intended to capture the intuition that an observer who, perhaps in order to ease his calculations, chooses to change his coordinates, should still be able to conduct his investigation within a classical

(commutative) context. The following result shows that this is indeed possible:

**Theorem 11** *$C(\mathcal{W})$ is diffeomorphism invariant.*

**Proof** Let $C$ be any context in $C(\mathcal{W})$. Then any Weyl operators $w_{d_1}^{S_1,\sigma_{S_1}}$ and $w_{d_2}^{S_2,\sigma_{S_2}}$ in $C$ commute:

$$w_{d_1}^{S_1,\sigma_{S_1}} \circ w_{d_2}^{S_2,\sigma_{S_2}} = w_{d_2}^{S_2,\sigma_{S_2}} \circ w_{d_1}^{S_1,\sigma_{S_1}}.$$

For any diffeomorphism $\phi : \Sigma \to \Sigma$ and $f \in C(\overline{\mathcal{A}})$, we have

$$A_\phi(w_d^{S,\sigma_S})(f) = w_{\phi(d)}^{\phi(S),\phi(\sigma_S)}(f) = f \circ \Theta_{\phi(d)}^{\phi(S),\phi(\sigma_S)}.$$

Let $\gamma$ be a path which starts from $\phi(S)$ without returning. (The proof for the remaining choices of $\gamma$ is similar.) By the same steps as in in lemma 5, it follows that (for any generalized connection $\overline{A} \in \overline{\mathcal{A}}$)

$$h_{\Theta_{\phi(d)}^{\phi(S),\phi(\sigma_S)}(\overline{A})}(\gamma) = d(\phi^{-1}(\gamma(0))\, h_{\overline{A}}(\gamma) = d(\phi^{-1}(\gamma(0)) \cdot h_{\phi_{\overline{\mathcal{A}}}^{-1}(\overline{A})}(\phi^{-1} \circ \gamma) = h_{\phi_{\overline{\mathcal{A}}}\left[\Theta_d^{S,\sigma_S}\left[\phi_{\overline{\mathcal{A}}}^{-1}(\overline{A})\right]\right]}(\gamma).$$

This establishes that

$$\Theta_{\phi(d)}^{\phi(S),\phi(\sigma_S)} = \phi_{\overline{\mathcal{A}}} \circ \Theta_d^{S,\sigma_S} \circ \phi_{\overline{\mathcal{A}}}^{-1}. \quad (*)$$

Using (*) and the definition $\alpha_\phi(f) \equiv f \circ \phi_{\overline{\mathcal{A}}}^{-1}$ repeatedly, we may now reason in the following manner:

$$\{\alpha_\phi \circ w_d^{S,\sigma_S} \circ \alpha_\phi^{-1}\}(f) = \alpha_\phi(w_d^{S,\sigma_S}(\alpha_\phi^{-1}(f))) = \alpha_\phi(\{\alpha_\phi^{-1}(f)\} \circ \Theta_d^{S,\sigma_S}) = \alpha_\phi(\{\alpha_\phi^{-1}(f)\} \circ (\phi_{\overline{\mathcal{A}}}^{-1} \circ \Theta_{\phi(d)}^{\phi(S),\phi(\sigma_S)} \circ \phi_{\overline{\mathcal{A}}})) = (\{\alpha_\phi^{-1}(f)\} \circ (\phi_{\overline{\mathcal{A}}}^{-1} \circ \Theta_{\phi(d)}^{\phi(S),\phi(\sigma_S)} \circ \phi_{\overline{\mathcal{A}}})) \circ \phi_{\overline{\mathcal{A}}}^{-1} = \{\alpha_\phi^{-1}(f)\} \circ \phi_{\overline{\mathcal{A}}}^{-1} \circ \Theta_{\phi(d)}^{\phi(S),\phi(\sigma_S)}$$
$$= \alpha_\phi(\alpha_\phi^{-1}(f)) \circ \Theta_{\phi(d)}^{\phi(S),\phi(\sigma_S)} = f \circ \Theta_{\phi(d)}^{\phi(S),\phi(\sigma_S)}.$$

This proves that (cf. proposition 3.34 in [9])

$$A_\phi(w_d^{S,\sigma_S}) = \alpha_\phi \circ w_d^{S,\sigma_S} \circ \alpha_\phi^{-1}.$$

It then follows by trivial steps that $A_\phi$ preserves commutativity of Weyl operators:

$$A_\phi(w_{d_1}^{S_1,\sigma_{S_1}}) \circ A_\phi(w_{d_2}^{S_2,\sigma_{S_2}}) = (\alpha_\phi \circ w_{d_1}^{S_1,\sigma_{S_1}} \circ \alpha_\phi^{-1}) \circ (\alpha_\phi \circ w_{d_2}^{S_2,\sigma_{S_2}} \circ \alpha_\phi^{-1}) = \alpha_\phi \circ (w_{d_1}^{S_1,\sigma_{S_1}} \circ w_{d_2}^{S_2,\sigma_{S_2}})$$
$$\circ \alpha_\phi^{-1} = \alpha_\phi \circ (w_{d_2}^{S_2,\sigma_{S_2}} \circ w_{d_1}^{S_1,\sigma_{S_1}}) \circ \alpha_\phi^{-1} = (\alpha_\phi \circ w_{d_2}^{S_2,\sigma_{S_2}} \circ \alpha_\phi^{-1}) \circ (\alpha_\phi \circ w_{d_1}^{S_1,\sigma_{S_1}} \circ \alpha_\phi^{-1}) =$$
$$A_\phi(w_{d_2}^{S_2,\sigma_{S_2}}) \circ A_\phi(w_{d_1}^{S_1,\sigma_{S_1}}).$$

The two remaining cases, involving the multiplicative operators by themselves and the mixed case of multiplicative operators and Weyl operators, are even simpler. □

Let us, briefly, consider the corresponding definition and result for gauge invariance of $C(\mathcal{W})$. Following, in part, Fleischhack ([9], def. 3.26), we define the *generalized gauge transformations* $\mathcal{G}$ as the set of maps $g : \Sigma \to SO(3)$. For each $g$, we say that $\beta_g(f)(\overline{A}) := f$ ($\overline{A}_g$), where $\overline{A}_g$ is given by $h_{\overline{A}_g}\gamma := g(\gamma(0))^{-1} h_{\overline{A}}\gamma \, g(\gamma(1))$. Then $(B_g(T_f))(\overline{A}) := T_f(\overline{A}_g)$ and $B_g(w_d^{S,\sigma_S}) := w_{g \cdot d \cdot g^{-1}}^{S,\sigma_S}$ define the transformations corresponding to (21) and (22) above.

**Definition 11** *A context* $C \in C(\mathcal{W})$ *is* gauge invariant *if, for any gauge transformation g : $\Sigma \to SO(3)$, the algebra generated by the image $B_g(C)$ is a commutative subalgebra in $C(\mathcal{W})$. If this holds for all contexts C, we say that $C(\mathcal{W})$ itself is* gauge invariant.

**Theorem 12** *$C(\mathcal{W})$ is gauge invariant.*

**Proof** Similar to th. 11. □

# 6. Conclusion

The approach to quantum gravity outlined above has been strictly limited to a globally hyperbolic space-time $\mathcal{M} \simeq \mathbb{R} \times \Sigma$. This foliation into separate entities "time" and "space" is dependent on the choice of an observer, and the requirement of 4-dimensional diffeomorphism invariance in general relativity is not fulfilled. Thus, the diffeomorphism invariance established in th. 12 does not hold for the general case, but solely for the restricted group of diffeomorphisms on the 3-dimensional hypersurface $\Sigma$. In a fully coordinate-free description of the laws of physics, we should expect the observational contexts in $C(\mathcal{W})$ to respect the full diffeomorphism group. This would dispel the implicit notion of a "meta-observer", slicing space-time from some arbitrary perspective. More precisely, in a complete topos theory on gravity, the unrealistic kinematical observational contexts would be replaced by dynamical contexts with physically realistic observables. This is an important task, but we will not enter into it here.

In the long run, the attempt should be made to sort out if a non-standard topos (that is, a topos different from **Sets**) is the most natural setting for LQG. Recall that, in LQG, the discrete nature of space-time emerges as a calculation within the theory. One may hope that, if the theory is stripped of non-physical content in the manner suggested by the topos approach, the auxilliary apparatus of standard differential geometry may be overcome, and the radical geometric structure of the theory may be founded on sound empiricist principles. It is unclear to what extent this program may be carried out. A less ambitious task is to work out toy examples to show what the physics of quantum gravity looks like from a stance within the topos $\tau_\mathcal{W}$ above.